# A New Hybrid Method of IPv6 Addressing in the Internet of Things

Nasrin Sadat Zarif
Department of
Computer Engineering,
Faculty of 17 Shahrivar,
Alborz Branch,
Technical and
Vocational University
(TVU), Alborz, Iran.
n.zariif@gmail.com

Hamed Najafi
Department of Electrical and
Computer Engineering, K.N.
Toosi University,
Tehran, Iran.
hamed.najafi@email.kntu.ac.ir

Mehdi Imani
Department of Computer
Engineering, Faculty of
Shahid Beheshti, Alborz
Branch, Technical and
Vocational University
(TVU), Alborz, Iran.
m.imani@gmail.com

Abolfazl Qiyasi Moghadam
Department of Electrical and
Computer Engineering, Faculty
of Shahid Shamsipour, Tehran
Branch, Technical and
Vocational University (TVU),
Tehran, Iran.
a.qiyasimoghadam@gmail.com

*Abstract*—Humans have always been seeking greater control over their surrounding objects. Today, with the help of the Internet of Things (IoT), we can fulfil this goal. In order for objects to be connected to the Internet, they should have an address, so that they can be detected and tracked. Since the number of these objects are very large and never stop growing, addressing space should be used, which can respond to this number of objects. In this regard, the best option is IPv6. Addressing has different methods, the most important of them introduced in this paper. The method presented in this paper is a hybrid addressing method which uses EPC and ONS IP. The method proposed in this paper provides a unique and hierarchical IPv6 address for each object. This method is simple and does not require additional hardware for implantation. Further, the addressing time of this method is short while its scalability is high, and is compatible with different EPC standards.

*Index Terms*—Addressing Methods, EPC, Internet of Things, IoT, IPv6

## I. Introduction

Although detection technology through radio frequency is not a new subject, this technology is less than a century has been used under the title of RFID. In 1980-1990, innovation in radio frequency-based technologies caused passive tags to be introduced into the market as part of this technology to achieve sufficient range across various uses [1].

As RFID grows day by day and the number of devices connected to the Internet increase worldwide, an interoperable standard was needed [2]. Therefore, the Electronic Product Code (EPC) was developed to resolve the problem of object detection by the MIT Auto-ID Center [3]. RFID is one of the widely used technologies in the Internet of Things (IoT), and it is expected that the integration of this technology with EPC would cause the development of IoT. Also, IoT plays an important role in both transmission level and distribution level such as microgrids, smart cities and Industry [4]. Fig. 1 represents the number of devices connected to IoT in 2018, according to the IHS report [5].

IoT technology plays a significant role in establishing communication between objects by connecting them through EPC by IPv6. Among the various technologies that have been used so far in this area are: RFID (Radio Frequency Identification) tags, NFC (Near Field Communication), mobile phones, Sensors, Low Power Wireless Personal Area Networks (6LoWPAN), IEEE 802.15.1 (Bluetooth), Machine to Machine (M2M), IEEE 802.15.4 (ZigBee) and etc.) [6,7,8].

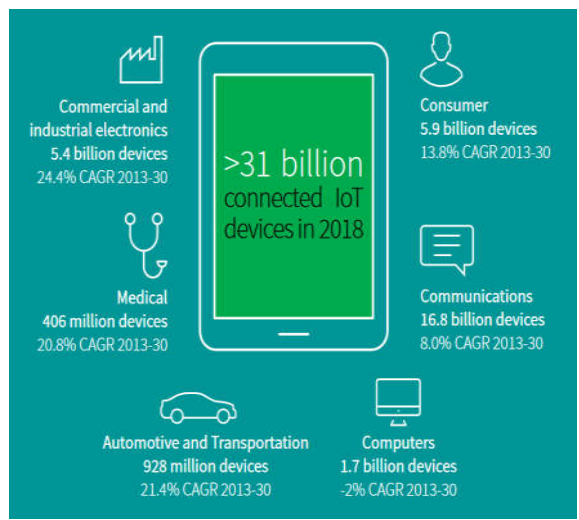

Figure. 1. The Number of Connected IoT Devices in [5].

RFID tags save an EPC code on their chip. The antenna then transfers this code to the reader. The reader reads the code and transfers to the Object Name System (ONS), which eventually can obtain precise information about the place of objects. However, to achieve information in a real-time fashion, the connection between the object should be established through the Internet by IPv6.

RFID tags cannot directly use IPv6. Thus, IPv6 is defined in different ways for the tag. Tag Reader has an ID which distinguishes the tagged object from other objects. Thus, ID can be used for creating and allocating an IPv6 address. This should be done in a simple and inexpensive way so that it can resolve the challenge of linking tags to objects on the Internet [9]. Various methods have been presented for the allocation of IPv6 to RFID tags, which are discussed in subsequent sections.

This paper has been organized as follows: Different EPC standards are examined in Section II. Then, Section III presents related works. Our proposed mechanism is then explained in Section IV. Eventually, our study is concluded in Section V.

## II. EPC STANDARD

With the establishment of EPC global organization, an intermediate means was developed which specifies which information can be collected and applied. Generally, the global EPC, as an intermediate, provides commercial benefits and great security for the users.

Currently, EPC exists in 64, 96, 128, 256 bits, and etc. sizes. The 96-bit codes are among the most common EPCs, and since one can detect an infinite number of objects by 96 bits (up to trillion), thus 96 bits can be adequate. EPC has Pure Identity EPC URI and EPC Tag URI forms, which generally have four sections: Header determines the EPC format; Manager Number specifies the object's manufacturer company; Object Class represents which type of objects use this EPC; and eventually, the serial number determines the unique code of that object. Fig. 2 demonstrates these components. Also, there are many EPC schemes which have different encodings as [10] mentions all of them. Each scheme has a specific application such as SGTIN is for general trade item, GIAI is for a fixed asset, SGLN is for location, USDOD is for US Dept. of Defense supply chain and etc.

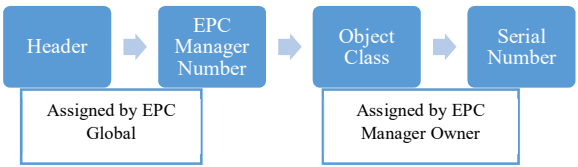

Figure. 2. The structure of EPC general form [10].

## III. RELATED WORKS

A very comprehensive study has been conducted on methods of addressing objects in IoT in [11]. The researchers who wish to further study this subject are suggested to refer to this paper.

IPv6 has a high capacity in addressing and can connect an infinite number of devices to the Internet [12]. The structure of IPv6 as defined in [13], is in the form of $2^{128}$ bits, where $2^{64}$ bits are allocated to the network address, and the other $2^{64}$ bits are assigned to the Host ID.

A middleware system has been proposed in [3] between reader and RFID applications based on EPC mapping. This middleware stores the server information and its database. The presented method uses the combination of 64 bits of network prefix with 64 bits of EPC (instead of EUI).

*Remarks:* This method is hierarchical and increases the interaction of the internet and RFID tag. But the performance of the presented method will decrease when faced with long EPCs.

The addressing method has been presented based on EPC using the XOR operator according to [14]. In this method, after converting to binary, EPC will have one the three states including less than 64 bits, equal to 64 bits, and greater than 64 bits. In the first state, a number of zeros (0) are added to the left side (left zero paddings). In the second state, the same binary number is directly used for the next step. Finally, in the third state, hash and compression functions are used to obtain the 64 bits. Then, the XOR operator is used to alter the bits obtained from the previous states. Eventually, in the last step, an IPv6 address is generated.

*Remarks:* This method is useful for simple and inexpensive implementation. Nevertheless, its problem is that if it uses compression and hash functions, temporal complexity grows as well as mechanism performance declines.

A method similar to [14] has been presented in [15]. This method uses EPC and performs all steps of the previous method, and the only difference is that instead of XOR operator, it uses OR (+) operator and after computations, it generates IPv6 address.

*Remarks:* This method has the advantage of being simple. On the other hand, it employs hash and compression functions for long EPCs, thus increasing temporal complexity. Further, this method can only be implemented for ID-based objects.

The Proposed method in [16] is an addressing method based on Cryptographically Generated Addresses(CGAs). This mechanism uses 64 bits of EPC (as Host ID) with 64 bits of Net ID for generating IPv6 addresses. It also uses compression strategies for long EPCs which increase computational overhead. This method uses EM4100 tags for implementation scenario.

*Remarks:* This method supports RFID technology for addressing, and there is no need for additional hardware for implementation, and it is also a hierarchical method.

An EPC mapping technique is presented in [17] for identifying home appliances using IoT. In order to translate the mapped EPC to IPv6 in this method, communication with devices takes place through codes (e.g. XML) via sensors (e.g. ZigBee).

*Remarks:* The advantage of This method is circuits communicate with users through mobile phone software interfaces and give the necessary information from the environment to the users. Consequently, the performance of this method is low in large areas and also it is not usable in heterogeneous environments.

The RFID Agent (RA) is presented in [18]. RA performs two operations which are storing information and sending them to the DHCP. The Net ID is generated by DHCP and the Host ID generated by RA. The combination of these two creates a new IPv6 address. RA and the home agent are responsible for managing the tags and storing their location.

*Remarks:* Although, this method provides the mobility feature for tags. Nonetheless, it is not usable for None-ID objects and therefore, the scalability of this method is low.

In [19], RFID & Mobile integration is provided to reduce transmission and conversion of data to multiple servers. This mechanism combines the 64 bits of network with 64 bits of EPC to construct a unique IPv6 address.

*Remarks:* In this method, a mobile phone is able to read the tag data and avoid the data transfer to servers which reduces both energy and time consumption. However, connecting the mobile phone to an unknown device requires authentication operation. Thus, it takes time as well as this mechanism is applicable only for mobile phones that support IPv6.

A simple yet practical method for EPC has been presented in [9]. In this method, the serial number part has been used in EPC and will be converted to binary. After this conversion, two states occur less than 64 bits and equal to 64 bits. In the first state, a number of ones (1) are added to the left side of the binary number (left one padding). In the second state, the same obtained binary numbers are used in subsequent steps of the mechanism without direct manipulation. Eventually, after being combined with net ID, an IPv6 address is generated.

*Remarks:* This method is simple and hierarchical and has low implementation costs and supports all EPC schemes for addressing. Further, unlike previous methods, due to not using hash and compression functions, it has less of temporal complexity, though it cannot be used for non-ID objects.

Another method has been presented in [20], which supports different standards such as ISO and EPC for addressing. The presented method communicates with RFID tags, reads the memory banks of the tag and applies the addressing mechanisms based on their values. Considering the addressing procedure, first, it recognizes the standard used in the tag, and if it is EPC standard, all EPC scheme is read and converted to binary. Different states occur at this step. Firstly, the obtained value might be less than 64 bits; in response, zero bits are added (left zero paddings). Secondly, the EPC value is equal to 64 bits, which is used for subsequent steps without any manipulation. And thirdly, EPC might be far greater than 64 bits. In this state, 64 bits of that are chosen and then combined with 64 bits of the reader net ID to generate a new IPv6 address. On the other hand, if the standard is ISO, then the entire ISO scheme is read, after which its serial number values are used. Eventually, by being combined with reader net ID, a new IPv6 address is obtained.

*Remarks:* The advantages of this method include supporting different standards for addressing, being hierarchical, simple implementation, no additional costs, and not using hash and compression functions, low temporal complexity and providing high scalability. However, in heterogeneous environments which have different sensors, it does not function well and does not support non-ID objects either.

## IV. PROPOSED METHOD

As shown in Fig. 3, in the mechanism proposed for generating a unique IPv6 address, EPC and ONS IP combination (high bits) which are connected to the reader of interest is used. EPC is presented in 64-bit, 96-bit, 128-bit, and 256-bit and other forms. Our proposed method covers all of its forms.

Concerning the addressing procedure, first EPC is converted to a binary form and in four-number groups. The obtained numbers are either less than or equal to 64 bits or are greater than 64 bits. The obtained number represents our bit of interest from ONS IP. For example, in an EPC with 45 bits, 128-45=83; we combine 83 bits of the high-order bits of IP ONS with 45 bits of the EPC code. As displayed in Fig. 3, in the EPC code, there is a part called the serial number, and another is called ID, which consists of a limited number of figures.

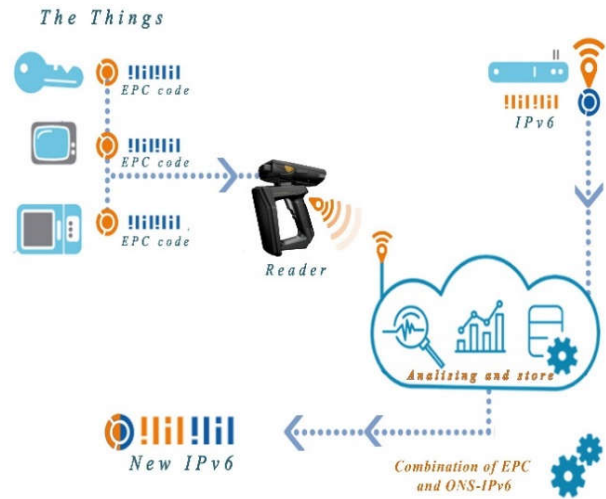

Figure. 3. Combination of EPC with ONS IPv6.

For EPC greater than 64 bits, we use ID part in URI or the numerical number of the EPC serial number. In this regard, the number of ID bits in the serial number is subtracted from 128 bits, and obtain the resulting number will be obtained from ONS IP (see Fig. 4, Fig.5, and Fig. 6). Algorithm 1 describes the proposed mechanism in detail.

$$128 - 40 = 88 \quad (1)$$

Equation (1) is the state that EPC is less than 64 bits. Where 128 bits are for ONS-IPv6, 40 bits belong to the EPC and 88 is the number of bits we should obtain for creating an IPv6 address. Fig. 4 shows the process of this case.

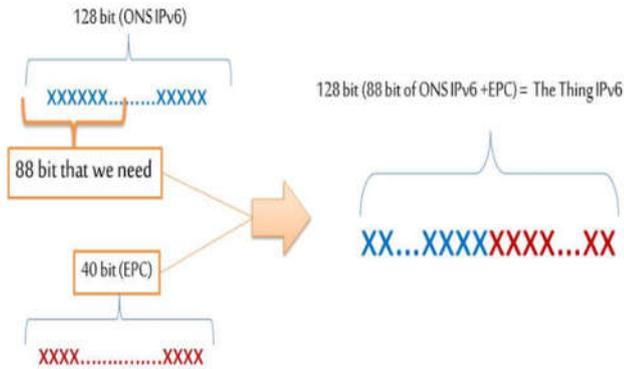

Figure. 4. Combination of EPC < 64 bit with ONS IP.

$$128 - 64 = 64 \quad (2)$$

Equation (2) is the state that EPC is equal to 64 bits. Where 128 bits are for ONS-IPv6, 64 bits belong to the EPC and 64 is the number of bits we should obtain for creating an IPv6 address. Fig. 5 shows the process of this case.

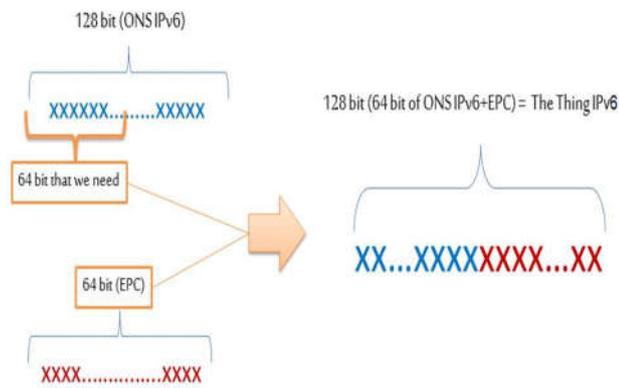

Figure. 5. Combination of EPC = 64 bit with ONS IP.

$$128 - 90 = 38 \quad (3)$$

Equation (3) is the state that EPC is more than 64 bits. If EPC has more than 64 bits, we use the numerical part of URI named ID in Serial Number (like Fig. 6). where 128 bits are for ONS-IPv6, 64 bits belong to the EPC Serial Number, and 64 is the number of bits we should obtain for creating an IPv6 address. Fig. 6 shows the process of this case.

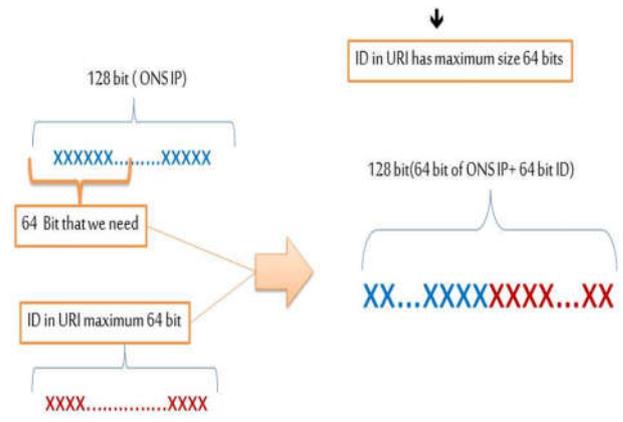

Figure. 6. Combination of EPC > 64 bit with ONS IP.

**Algorithm. 1.** The pseudo code of the proposed addressing method.

```
#Read the EPC by reader
#Read the serial number by reader
#get the ONS IPv6
#for example EPC is 32
EPC = 32
serial_number=0
shortage_of_bits=0
address_list = list()
for i in range(128)
    address_list[i] = ipv6 of Ons
need_list = list()
if EPC <=64:
shortage_of_bits = 128 – EPC
for i in range(shortage_of_bits):
   need_list[i] = adresslist[i]
new_ip =  need_list + EPC
else:
#serial_number = serial number of product
   shortage_of_bits = 128 - serial_number
for i in range(shortage_of_bits) :
   need_list[i] = adresslist[i]
new_ip = need_list + serial_number
```

Since the proposed mechanism does not need extra hardware, it will have a higher speed and far lower computational overhead compared to other methods. This method functions the same for EPCs with different number of bits, and at EPCs greater than 64 bits, it does not need compression functions. Fig. 7 provides the simulation results of the proposed approach in Python.

```
===========For EPC <= 64 Bits ============
EPC is :  961683854154598
ONS's IP is :
3ffe:ffff:4004:1952:0000:7251:bc9b:a73f
Unique New IPv6 :
3ffe:ffff:4004:1952:22:25c6:89d1:fb66
===========For EPC > 64 Bits ============
Serial number is :  37375918425780
ONS's IP is :
3ffe:ffff:4004:1952:0000:7251:bc9b:a73f
Unique New IPv6 is :
3ffe:ffff:4004:1952:0: 61fe:4257:46b4
```

Figure. 7.  Results of the proposed mechanism.

## V. CONCLUSION

The aim of presenting this paper has been connecting objects through the Internet, proper performance, and gaining better control over the surroundings. The proposed mechanism is hybrid addressing which can develop unique addresses for objects. This method does not need extra hardware, computational overhead, and complex operations for using EPC with a different number of bits in addressing objects. Its simple functioning contributes to saving costs and time, enabling us to address objects by EPC.